\documentclass[fleqn]{2020SCGE} 
\usepackage{siunitx}
\usepackage{hyperref}

\usepackage{booktabs}
\usepackage{color}
\usepackage{verbatim}
\usepackage{float}

\usepackage{graphicx}
\usepackage{amsmath}
\usepackage{amssymb}	
\usepackage{diagbox}
\usepackage{threeparttable}

\usepackage{ulem}
\usepackage[numbers]{natbib}
\usepackage{longtable, threeparttablex, booktabs, url}
\usepackage{multirow} 
\usepackage{array} 
\usepackage{makecell} 
\usepackage{colortbl}
\usepackage{xcolor}
\usepackage{tabularx} 

\usepackage{color}

\begin{document}
\ensubject{subject}
\ArticleType{Article}
\SpecialTopic{SPECIAL TOPIC: }
\Year{2026}
\Month{xxx}
\Vol{xx}
\No{x}
\DOI{xx}
\ArtNo{000000}
\ReceiveDate{xx}
\AcceptDate{xx}
\AuthorMark{J. Zhang}

\AuthorCitation{J. Zhang, et al.}
\title{Discovery of a millisecond pulsar with a CO white dwarf companion}

\author[1]{Jie Zhang}{{zhangjie_mail@126.com}} 
\author[2]{Ze-Rui Wang}{{zeruiwang@qlnu.edu.cn}}
\author[3,4]{Lei Zhang}{{leizhang996@nao.cas.cn}}
\author[3]{Yulan Liu}{}
\author[5]{Alessandro Ridolfi}{}
\author[6,7]{Meng Guo}{}
\author[8]{Di Li}{dili@tsinghua.edu.cn}
\author[9]{\\Ryan S. Lynch}{}
\author[2]{Cong Wang}{}
\author[3]{Pei Wang}{}
\author[7]{Mengmeng Ni}{}
\author[7]{Jiale Hu}{}
\author[1]{Mengquan Liu}{}
\author[2]{Zhie Liu}{}
\author[2]{\\Bo Han}{}
\author[2]{Chenchen Miao}{}

\address[1]{School of Arts and Sciences, Shanghai Dianji University,Shanghai {\rm 201306}, China}
\address[2]{College of Physics and Electronic Engineering, Qilu Normal University, Jinan {\rm 250200}, China}
\address[3]{State Key Laboratory of Radio Astronomy and Technology, National Astronomical Observatories, Chinese Academy of Sciences, Beijing {\rm 100101}, China}
\address[4]{Centre for Astrophysics and Supercomputing, Swinburne University of Technology, P.O. Box 218, Hawthorn, VIC {\rm 3122}, Australia}
\address[5]{Fakult\"at f\"ur Physik, Universit\"at Bielefeld, Postfach 100131, D-33501 Bielefeld, Germany}
\address[6]{National Supercomputing Center in Jinan, Qilu University of Technology, Jinan {\rm 250103}, China}
\address[7]{Jinan Institute of Supercomputing Technology, Jinan {\rm 250103}, China}
\address[8]{New Cornerstone Laboratory, Department of Astronomy, Tsinghua University, Beijing {\rm 100084}, China}
\address[9]{Green Bank Observatory, Green Bank, WV {\rm 24401}, USA}

\abstract{We report the discovery and characterization of PSR~J1810$-$0623, a fully recycled millisecond pulsar with a spin period of 4.55 ms, discovered with the Five-hundred-meter Aperture Spherical radio Telescope (FAST) and followed up with FAST and the Green Bank Telescope (GBT). A phase-connected timing solution spanning over 6.5 years reveals a 15.4-day binary orbit with extremely low eccentricity ($e \simeq 1.5 \times 10^{-5}$). Assuming a neutron-star mass of 1.4\,$M_\odot$, the inferred companion median mass ($\sim$0.64\,$M_\odot$) is consistent with a carbon–oxygen white dwarf, indicating an evolutionary origin in an intermediate-mass X-ray binary.  
The system’s properties closely resemble those of other massive white dwarf binaries thought to form via Case A Roche lobe overflow, suggesting a prolonged accretion phase during which the neutron star was efficiently recycled. 
Polarimetric analysis of FAST data yields a moderate degree of linear polarization and a rotation measure of $86.6 \pm 0.6$\,rad\,m$^{-2}$, offering constraints on the Galactic magnetic field. The inferred characteristic age ($\sim$32\,Gyr) and low surface magnetic field ($\sim$$10^8$\,G) indicate a highly recycled pulsar. Proper-motion measurements imply a modest transverse velocity, consistent with those of recycled millisecond pulsars in the Galactic field. Although the proximity of the globular cluster Pal~7 raises the possibility of a dynamical origin, discrepancies in dispersion measure and proper motion argue against a physical association. PSR~J1810$-$0623 adds to the rare class of long-orbital period MSP-CO WD systems and provides a valuable laboratory for studying pulsar recycling, binary evolution, and Galactic structure.
}
\maketitle

\begin{multicols}{2}
\section{Introduction} \label{sec:intro}
Millisecond pulsars (MSPs) are neutron stars that have been spun up to rapid rotation ($P \lesssim 30$ ms) and are characterized by low surface magnetic fields ($B_{\rm surf} \lesssim 10^{10}$ G)\footnote{This working definition of an MSP is simplified (c.f.~\cite{Lee2012}), but facilitates consistent comparisons in this work.}, through the accretion of mass and angular momentum from a binary companion~\citep{Alpar1982, Bhattacharya1991}. As precise cosmic clocks, MSPs serve as powerful tools for testing general relativity (e.g. \cite{Freire2024}), probing the neutron star equation of state through mass measurements (e.g. \cite{Demorest2010, Ozel2016, Lattimer2021}), tracing magnetic field evolution (e.g. \cite{Zhang2025}), and detecting nanohertz gravitational waves through pulsar timing arrays (e.g. \cite{Agazie2023, EPTA2023, Reardon2023, Xu2023, Miles2025}).

To date, over 800 MSPs have been discovered, the majority of which reside in binary systems with a wide range of companion types and orbital properties\footnote{Data are from the ATNF pulsar catalogue V2.7.0 \url{https://www.atnf.csiro.au/research/pulsar/psrcat/}\label{ATNF_catalog}}. Companion stars span neutron stars (NSs), white dwarfs (WDs), main sequence (MS) stars, and ultra-light (UL) objects such as brown dwarfs, ablated WD remnants, or even planetary-mass companions ($m_c \lesssim 0.1,M_\odot$), as summarized in \cite{Wang2025}. Among these, MSPs with carbon-oxygen white dwarf (CO-WD) companions are particularly interesting, as they are believed to result from intermediate-mass X-ray binary (IMXB) progenitors \cite{Tauris2012}. These systems generally experience shorter accretion phases and are expected to exhibit longer spin periods, wider orbits ($P_{\rm orb} \gtrsim 5$ days), and higher eccentricities compared to systems with helium WD (He-WD) companions. A landmark case is PSR~J1614$-$2230, which yielded the first $\sim$2\,$M_\odot$ neutron star mass via Shapiro delay \cite{Demorest2010}. Understanding the formation and evolution of MSP+CO-WD binary systems provides key constraints on evolutionary pathways, accretion efficiency, and post-accretion orbital dynamics.

In this paper, we report the discovery and phase-connected timing of PSR~J1810$-$0623, a 4.55-ms pulsar in a 15.4-day binary orbit with a low eccentricity ($e \sim 1.5 \times 10^{-5}$). The system was discovered using the Five-hundred-meter Aperture Spherical radio Telescope (FAST), and follow-up observations with both FAST and the Green Bank Telescope (GBT) have yielded a consistent orbital solution. Assuming a canonical neutron star mass of 1.4 $M_\odot$, the inferred median companion mass of $\sim$0.64 $M_\odot$ strongly supports the presence of a CO-WD companion.

This configuration places PSR~J1810$-$0623 among the rare but significant class of millisecond pulsar binaries with relatively massive white dwarf companions and unusually short spin periods. Similar to PSR~J1614$-$2230, it may provide an opportunity for future Shapiro delay measurements. PSR~J1810$-$0623 thus adds to the small but growing population of low-eccentricity, long orbital-period MSP+CO-WD systems, offering a valuable laboratory for exploring pulsar recycling and compact binary evolution.

\section{Observations} \label{sec:obs}
We have carried out observations of PSR~J1810$-$0623 with both the FAST and GBT radio telescopes. A summary of all the observations used in this paper is listed in Table~\ref{tab:J1810_obs}. In the following section, we describe the observations made with each telescope.
\begin{table*}
\caption{Summary of the FAST and GBT observations used in this work.}
\label{tab:J1810_obs}
\centering
\begin{tabular}{clcccccc}
\hline
Observation & Start time & Start time & Central frequency & Sampling time & Bandwidth & Number of & Observation\\
ID          &  (Date)    &  (MJD)     & (MHz)             & ($\mu s$)     &  (MHz)    & channels  & Length (minute)\\\hline
\multicolumn{8}{c}{\bf{FAST}}\\
01 & 2019 Jul 10 & 58674.63 & 1250 & 49.15 & 500 & 4096 & 1 \\
02 & 2023 Aug 11 & 60167.55 & 1250 & 49.15 & 500 & 4096 & 30\\
03 & 2024 Oct 01 & 60584.37 & 1250 & 49.15 & 500 & 4096 & 20\\
04 & 2025 Aug 11 & 60898.60 & 1250 & 49.15 & 500 & 4096 & 10\\
05 & 2025 Aug 16 & 60903.56 & 1250 & 49.15 & 500 & 4096 & 10\\
06 & 2025 Aug 22 & 60909.51 & 1250 & 49.15 & 500 & 4096 & 10\\
07 & 2025 Aug 28 & 60915.56 & 1250 & 49.15 & 500 & 4096 & 10\\
08 & 2025 Dec 03 & 61012.30 & 1250 & 49.15 & 500 & 4096 & 20\\
09 & 2025 Dec 10 & 61019.20 & 1250 & 49.15 & 500 & 4096 & 20\\
10 & 2025 Dec 17 & 61026.26 & 1250 & 49.15 & 500 & 4096 & 20\\
11 & 2025 Dec 24 & 61033.24 & 1250 & 49.15 & 500 & 4096 & 20\\
12 & 2026 Jan 01 & 61041.15 & 1250 & 49.15 & 500 & 4096 & 20\\
13 & 2026 Jan 08 & 61048.15 & 1250 & 49.15 & 500 & 4096 & 20\\
\multicolumn{8}{c}{\bf{GBT}}\\
01 & 2024 Apr 28 & 60427.22 & 1440 & 10.24 & 800 & 512 & 444\\
02 & 2024 May 03 & 60433.22 & 1440 & 10.24 & 800 & 512 & 444\\
03 & 2024 Dec 18 & 60662.58 & 1440 & 10.24 & 800 & 512 & 120\\
04 & 2024 Dec 24 & 60668.55 & 1440 & 10.24 & 800 & 512 & 132\\
05 & 2024 Dec 31 & 60675.55 & 1440 & 10.24 & 800 & 512 & 132\\
06 & 2025 Jan 14 & 60689.57 & 1440 & 10.24 & 800 & 512  & 132\\
\hline
\end{tabular}
\end{table*}

\subsection{FAST observations}
We conducted 13 observations of PSR~J1810$-$0623 with the FAST between 2019 July 10 and 2026 January 08, using the 19-beam receiver. The initial observation, performed under project 3045, utilized the FAST snapshot mode \cite{Han2021}, which consists of four one-minute pointings slightly offset from each other to ensure uniform sky coverage while minimizing telescope slewing. The central beam was pointed at $\alpha = 18^{\rm h} 11^{\rm m} 20^{\rm s}.6$, $\delta = -06^{\circ} 23^{'} 20^{''}.9$ (J2000).

Twelve follow-up observations were subsequently conducted using the central beam of the FAST 19-beam receiver (with a beamwidth of 3$^{'}$ at 1250 MHz) in tracking mode. Each session began with a one-minute noise diode injection to enable polarization calibration, although no flux calibrator observations were performed. All follow-up pointings targeted the same sky position as the original discovery beam, centered at $\alpha = 18^{\rm h} 10^{\rm m} 33^{\rm s}.1$, $\delta = -06^{\circ} 22^{'} 21^{''}.3$ (J2000), and were conducted under three observing programs: PT2024\_0233, PT2025\_0186, and DDT2025\_6.

All FAST observations were conducted in pulsar search mode, recording full-Stokes parameters with 8-bit sampling and a time resolution of 49 $\mu$s. The recorded frequency band spanned 1000–1500 MHz, split into 4096 channels and centered at 1250 MHz. Due to bandpass roll-off, the effective usable bandwidth was limited to 1050–1450 MHz.

\subsection{GBT observations}
To further characterize PSR~J1810$-$0623, we conducted two long-duration follow-up observations with GBT, each lasting 7.5 hours, on 2024 April 28 and May 3 (project GBT-24A-465). These were supplemented by four additional 2-hour observations, spaced approximately one week apart, between 2024 December 18 and 2025 January 14 (project GBT-24B-505). While full-Stokes polarization data—including self and cross terms—were recorded, the analysis presented here uses only the total intensity, obtained by summing over all polarizations and only with total intensity data.

All GBT observations were performed with the L-band receiver centered at 1440 MHz, with a nominal total bandwidth of 800 MHz divided into 512 frequency channels. Data were sampled with 8-bit precision at a time resolution of 10.24 $\mu$s. Real-time coherent de-dispersion was applied using a dispersion measure (DM) of 24.4 pc cm$^{-3}$, derived from the initial FAST discovery.

\section{Analyses and Results} \label{sec:res}
\begin{table*}
\centering
\caption{Parameters for PSR~J1810$-$0622}
\label{tb:J1810_par}
\renewcommand{\arraystretch}{1.0}
\setlength{\tabcolsep}{8mm}{
\begin{tabular}{lllll}
\hline
\multicolumn{3}{l}{Parameter}                             & \multicolumn{2}{c}{Value} \\ \hline

\multicolumn{5}{c}{Data reduction parameters}\\ \hline        
\multicolumn{3}{l}{MJD range of FAST}                      & \multicolumn{2}{c}{58674.63 to 61048.15 }\\
\multicolumn{3}{l}{MJD range of GBT}                       & \multicolumn{2}{c}{60427.22 to 60689.57}\\
\multicolumn{3}{l}{Reference epoch (MJD)}                  & \multicolumn{2}{c}{60433.328431}\\
\multicolumn{3}{l}{Number of ToAs}                         & \multicolumn{2}{c}{140}\\
\multicolumn{3}{l}{Residuals rms ($\mu$s)}                 & \multicolumn{2}{c}{4.93}\\
\multicolumn{3}{l}{Reduced $\chi^2$}                       & \multicolumn{2}{c}{1.009}\\
\multicolumn{3}{l}{Solar system ephemeris}                 & \multicolumn{2}{c}{440}\\
\multicolumn{3}{l}{Binary model}                           & \multicolumn{2}{c}{ELL1}\\
\hline 

\multicolumn{5}{c}{Measured parameters}\\ \hline        
\multicolumn{3}{l}{R.A., $\alpha$ (J2000)}           & \multicolumn{2}{c}{$18^{\rm h}\;10^{\rm m}\;30^{\rm s}.3585(2)$}\\   
\multicolumn{3}{l}{Dec., $\delta$ (J2000)}           & \multicolumn{2}{c}{$-06^{\circ}\;23^{'}\;08^{''}.015(9) $}\\
\multicolumn{3}{l}{Proper motion in $\alpha$, $\mu_{\alpha}$ (mas yr$^{-1}$)}   & \multicolumn{2}{c}{4.7(1.6)}\\
\multicolumn{3}{l}{Proper motion in $\delta$, $\mu_{\delta}$ (mas yr$^{-1}$)}   & \multicolumn{2}{c}{$-$16(8)}\\
\multicolumn{3}{l}{Spin frequency, $f$ (s$^{-1}$)}                              & \multicolumn{2}{c}{219.70094829921(4)}\\
\multicolumn{3}{l}{Spin frequency first time derivative, $\dot{f}$ (s$^{-2}$)}  & \multicolumn{2}{c}{$-$1.099(4) $\times$ 10$^{-16}$}\\
\multicolumn{3}{l}{Orbital period, $P_{b}$ (days)}                   & \multicolumn{2}{c}{15.37583407(2)}\\
\multicolumn{3}{l}{Projected semimajor axis, $\chi$ (lt-s)}          & \multicolumn{2}{c}{20.7854562(8)}\\
\multicolumn{3}{l}{Epoch of the ascending node, $T_{\rm asc}$ (MJD)}     & \multicolumn{2}{c}{60416.0178494(5)}\\
\multicolumn{3}{l}{$e$ sin$\omega$, $\varepsilon_{1}$}               & \multicolumn{2}{c}{$-$1.260(16) $\times$ 10$^{-5}$}\\
\multicolumn{3}{l}{$e$ cos$\omega$, $\varepsilon_{2}$}               & \multicolumn{2}{c}{0.900(16) $\times$ 10$^{-5}$}\\
\multicolumn{3}{l}{Dispersion measure, DM (pc cm$^{-3}$)}         & \multicolumn{2}{c}{24.402(2)}\\
\multicolumn{3}{l}{Rotation Measure$^{a}$, RM (rad m$^{-2}$)}     & \multicolumn{2}{c}{86.6(6)}\\
\multicolumn{3}{l}{Flux density at 1.25 GHz$^{a}$ ($\mu$Jy)}               & \multicolumn{2}{c}{73(20)}\\ 
\multicolumn{3}{l}{Pulse width at 50\% of peak$^{a}$, $W_{50}$ ($^{\circ}$)}   & \multicolumn{2}{c}{95.17}\\
\multicolumn{3}{l}{Pulse width at 10\% of peak$^{a}$, $W_{10}$ ($^{\circ}$)}   & \multicolumn{2}{c}{244.32}\\
\multicolumn{3}{l}{Percentage linear polarization$^{a}$, $L/I$ (\%)}                & \multicolumn{2}{c}{31.0(6)}\\
\multicolumn{3}{l}{Percentage circular polarization$^{a}$, $V/I$ (\%)}              & \multicolumn{2}{c}{1.0(5)}\\
\multicolumn{3}{l}{Percentage absolute circular polarization$^{a}$, $|V|/I$ (\%)}   & \multicolumn{2}{c}{7.9(5)}\\
\hline  

\multicolumn{5}{c}{Derived Parameters}\\ \hline
\multicolumn{3}{l}{Galactic longitude, $l$ (deg)}                              & \multicolumn{2}{c}{$22.53$}\\ 
\multicolumn{3}{l}{Galactic longitude, $b$ (deg)}                              & \multicolumn{2}{c}{$6.11$}\\ 
\multicolumn{3}{l}{Estimated distance$^{b}$ (pc)}                              & \multicolumn{2}{c}{207, 1024}\\ 
\multicolumn{3}{l}{Spin period, $P$ (ms)}                                      & \multicolumn{2}{c}{4.5516417099761(8)}\\
\multicolumn{3}{l}{Spin period first time derivative, $\dot{P}$ ( s s$^{-1}$)}  & \multicolumn{2}{c}{2.276(8)$\times$10$^{-21}$}\\
\multicolumn{3}{l}{Surface dipole magnetic field, $B_{\rm s}$ (G)}     & \multicolumn{2}{c}{$1.03\times10^{8}$}\\ 
\multicolumn{3}{l}{Spin-down luminosity, $\dot{E}$ (erg s$^{-1}$)} & \multicolumn{2}{c}{$9.53\times10^{32}$}\\ 
\multicolumn{3}{l}{Characteristic age, $\tau_{\rm c}$ (Gyr)}           & \multicolumn{2}{c}{31.68}\\ 
\multicolumn{3}{l}{Eccentricity, $e$}                         & \multicolumn{2}{c}{1.548(16) $\times$ 10$^{-5}$}\\
\multicolumn{3}{l}{Longitude of periastron, $\omega$ (deg)}   & \multicolumn{2}{c}{305.5(6)}\\
\multicolumn{3}{l}{Mass function$^{b}$, $f (M_{\odot})$}      & \multicolumn{2}{c}{0.040783460(4)}\\ 
\multicolumn{3}{l}{Minimum companion mass$^{b}$, $M_{\rm c,min} (M_{\odot})$}  & \multicolumn{2}{c}{0.5344}\\ 
\multicolumn{3}{l}{Median companion mass$^{b}$, $M_{\rm c,med} (M_{\odot})$}   & \multicolumn{2}{c}{0.6391}\\ 
\hline
\multicolumn{5}{l}{{\bf Notes.} Numbers in parentheses represent uncertainties on the last digit.}\\
\multicolumn{5}{l}{$^{a}$ Parameter derived from FAST data only.}\\
\multicolumn{5}{l}{$^{b}$ Parameter derived from the YMW16~\cite{Yao2017} and NE2001~\cite{Cordes2002} model.}\\
\multicolumn{4}{l}{$^{c}$ The companion masses assume a pulsar mass of 1.4 $M_{\odot}$. The minimum and median}\\
\multicolumn{4}{l}{~~~masses assume an inclination angle of 90$^{\circ}$ and 60$^{\circ}$, respectively.}
\end{tabular}}
\end{table*}

\subsection{Discovery of PSR~J1810$-$0623}
We searched for new pulsars in the FAST observations using a pipeline based on the \texttt{PRESTO} software suite~\cite{Ransom2002}. The raw data were first cleaned of radio frequency interference (RFI) using the \texttt{rfifind} routine and subsequently de-dispersed over a dispersion measure (DM) range of 0–500 pc cm$^{-3}$. A Fourier-domain search was then performed on the de-dispersed time series using \texttt{accelsearch}, with a maximum Fourier drift rate of $z_{\rm max} = 50$, enhancing sensitivity to pulsars in compact binary systems.
In parallel, we conducted single-pulse searches using the \texttt{single\_pulse\_search.py} routine, identifying candidates with signal-to-noise ratios (S/N) greater than 8. The routine applied a range of boxcar filter widths from 1 to 100 samples to each de-dispersed time series to detect transient signals.
All periodic and transient candidates were visually inspected after removing narrowband and impulsive RFI. This manual inspection was facilitated through an online web platform\footnote{\url{https://starfinity.at.suan.wang}}, developed by the team at the Jinan Institute of Supercomputing Technology.

Through the methods described above, we discovered an MSP candidate with a spin period of 4.55 ms and a DM of 24.4 pc cm$^{-3}$ on the one-minute test observation made on 2019 July 10 (see Section \ref{sec:obs}). This candidate was later confirmed as a new MSP, named PSR~J1810$-$0623, through the 30-minute follow-up observation conducted with FAST on 2023 August 11. Across this epoch, we observed varying line-of-sight accelerations, providing clear evidence of binary motion. 

\subsection{Timing Analysis}
To constrain the orbital parameters and further characterize PSR~J1810$-$0623, we conducted an additional 18 follow-up observations with FAST and GBT at L-band (Table~\ref{tab:J1810_obs}) over 6.5 yr. First, we used the six long GBT observations to re-detect the pulsar and measure the apparent spin period, $P_{\rm obs}$, and spin period derivative, $\dot{P}_{\rm obs}$, in each epoch to high significance\footnote{The FAST pointings were excluded because they are too short to be able to measure the spin period derivative within each individual observation to high significance}. The ($P_{\rm obs}, \dot{P}_{\rm obs}$) measurements were used to infer the rough orbital parameters using the Period-Acceleration diagram (see \citep{Freire+2001} and its Erratum \citep{Freire+2009}) method. This method revealed a circular pulsar orbit, with a period $P_{\rm b} \approx 15.4~{\rm days}$ and a projected semi-major axis of $x_{\rm p} \approx 20.8$\, lt-s. These values were used as input parameters for the \texttt{PRESTO}'s \texttt{fit\_circular\_orbit.py} script, which fits the variations of $P_{\rm obs}$ and $\dot{P}_{\rm obs}$ as a function of time. The script returned more precise values of $P$,  $P_{\rm b}$ and $x_{\rm p}$, which were in turn used to build a first ephemeris for the pulsar.

\begin{figure*}[]
    \centering
    \includegraphics[width=0.9\linewidth]{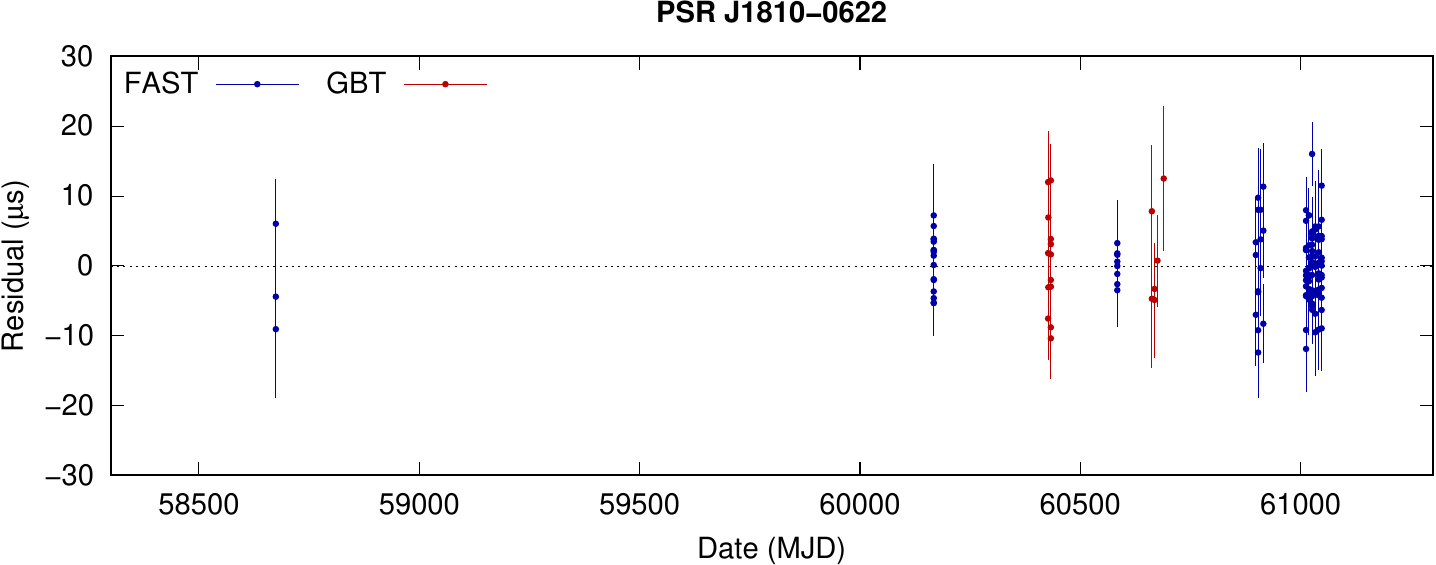}
    \vskip 1 cm
    \includegraphics[width=0.9\linewidth]{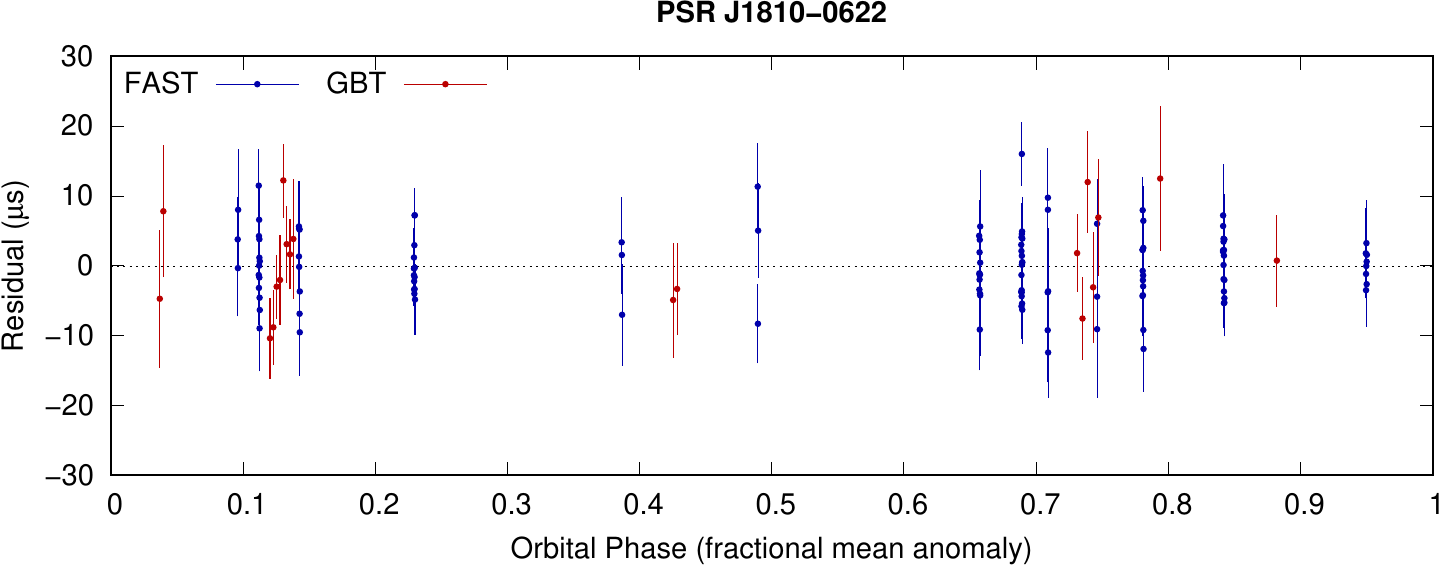}
    \caption{Timing residuals as a function of time (top panel) and orbital phase (bottom panel). The blue points indicate FAST ToAs and the red points are GBT ToAs.}
    \label{fig:J1810_res}
\end{figure*}

This ephemeris was used with the \textsc{DSPSR}\footnote{\url{https://dspsr.sourceforge.net/}}~\cite{vanStraten2011} software to re-fold all the FAST and GBT observations of PSR~J1810$-$0623, using 30-second-long subintegrations and 256 profile bins. The pulsar was folded correctly in all the observations, confirming the goodness of our initial orbital model. For each archive, we manually removed radio frequency interference in both frequency and time domains, and later summed groups of frequency channels and sub-integrations so as to have sufficiently high-signal-to-noise-ratio integrated pulse profiles. We then cross-correlated them with a profile template using the \texttt{pat} routine from the \textsc{PSRCHIVE}\footnote{\url{https://psrchive.sourceforge.net/}}~\cite{Hotan2004} package, and extracted 132 pulse times of arrival (ToAs). These were given, together with the starting ephemeris, to the \textsc{DRACULA} \footnote{\url{https://github.com/pfreire163/Dracula}}~\cite{Freire2018} software. \textsc{DRACULA} uses the \textsc{TEMPO}\footnote{\url{http://tempo.sourceforge.net}} timing software under the hood, to cleverly determine the exact rotation count of the NS between each observation (with the exception of a single arbitrary offset that we allowed between the FAST and GBT datasets to account for instrumental timing offsets between the two telescopes). Using the ELL1 binary model \citep{Lange+2001} and ignoring clock corrections (setting the CLK key to ``UNCORR'' in the ephemeris) \textsc{DRACULA} was able to find a single, unambigous, phase-connected solution, although the associated reduced chi square was somewhat high, 2.36. In fact, the timing residuals showed some small wavy trends, hinting to a possible unaccounted proper motion. For this reason, we used the solution found to re-fold the search-mode data, re-extract the ToAs (140, this time), and carry out a second, more refined, iteration of timing analysis with  \textsc{DRACULA}. In this new iteration, we enabled the fitting of the proper motion along both coordinate. Moreover, we enabled the use of the UTC(NIST) time scale, and included the most up-to-date (updated to 2026 January 27) clock corrections files available for the FAST\footnote{\url{https://github.com/NAOC-pulsar/FAST_ClockFile}} and GBT\footnote{\url{https://raw.githubusercontent.com/ipta/pulsar-clock-corrections/main/tempo/clock/time_gbt.dat}} telescopes, as well as for the UT1 time standard\footnote{\url{https://raw.githubusercontent.com/ipta/pulsar-clock-corrections/refs/heads/main/tempo/clock/ut1.dat}}. Also, the ToAs from the GBT and FAST datasets (between which we kept allowing an arbitrary offset) were obtained cross-correlating two different standard profiles, constructed from their respective observations. This allowed us to gain slightly better uncertainties in the ToAs. 
This time, \textsc{DRACULA} found three possible solutions, with a reduced $\chi^2$ of 1.009, 1.458 and 1.78, respectively. Besides having a much higher reduced $\chi^2$, the latter two solutions returned a measured total proper motion of 425 and 459 mas~yr$^{-1}$ respectively, larger than the largest precisely measured proper motion for any pulsar ($\approx 373.9$~mas yr$^{-1}$ in B1133+16). On the other hand, the first solution gave a much better fit, and a much more realistic total proper motion of $\approx 16.8$~mas\,yr$^{-1}$. Therefore, we deem this solution as the correct one, and we report the relative parameters in Table \ref{tb:J1810_par}. The corresponding timing residuals are shown as a function of time and orbital phase in Figure~\ref{fig:J1810_res}.

\subsection{Polarization Profile and Flux Density}
To investigate the pulsed emission properties of PSR~J1810$+$0623, we constructed an average pulse profile by summing all FAST observations in time, using the phase-connected timing solution to ensure accurate phase alignment. Parallactic angle corrections were applied using the \texttt{pac} routine, and Stokes parameters were calibrated following the conventions described in \cite{vanS10}.

\begin{figure*}[]
    \centering
    \includegraphics[width=0.6\linewidth]{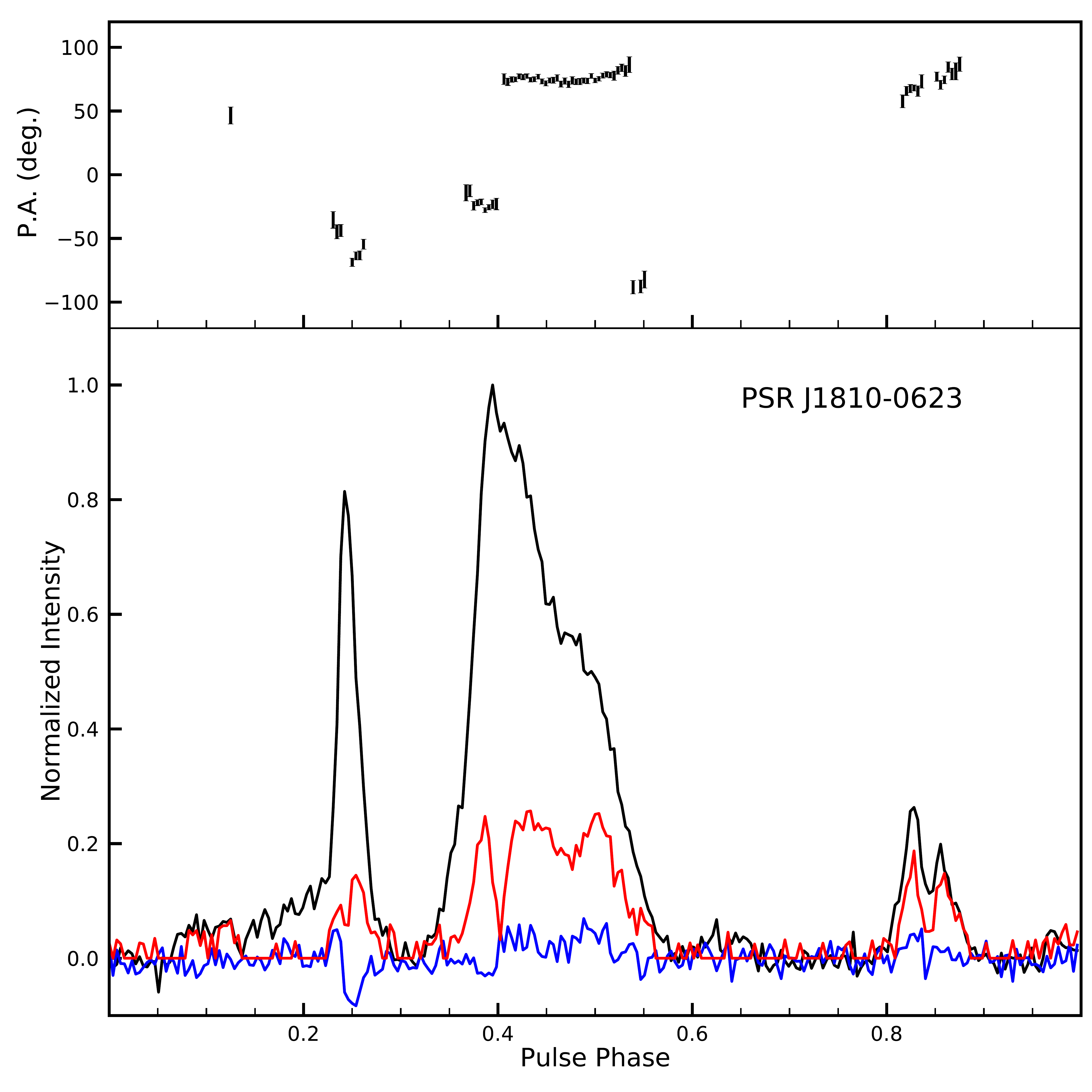}
    \caption{Integrated polarization profile of the PSR J1810-0623 from FAST data at 1250 MHz. The red line is the linear polarization profile, the blue line is the circular polarization profile, and the black line is the normalized total intensity profile. Black dots in the top panel give the linear position angle (PA) referred to the overall band center of the integrated profiles.}
    \label{fig:J1810_PolProf}
\end{figure*}

From the combined dataset, we measured a rotation measure (RM) of $86.6 \pm 0.6$ rad m$^{-2}$ using the \texttt{rmfit} tool in \textsc{psrchive}. Figure~\ref{fig:J1810_PolProf} shows the resulting average polarization profile, integrated across the full FAST observing band centered at 1250 MHz. The pulse widths at 50\% and 10\% of peak intensity ($W_{50}$ and $W_{10}$) were measured from noise-free templates using \texttt{paas} and visualized with \texttt{pdv}. The fractional linear polarization and position angle (PA) swing of the linear component were derived following the procedure outlined in \cite{Zhang2025}.

We estimate a mean flux density of $S_{\rm 1250MHz} \approx 73 \pm 20\ \mu$Jy, based on the radiometer equation~\cite{Lorimer04} and incorporating the known system parameters for FAST. The apparent variability in flux across observing epochs is consistent with modulation due to interstellar scintillation, as expected for pulsars at this DM~\cite{Gitika2023}. The measured values of $W_{50}$, $W_{10}$, and the flux density are listed in Table~\ref{tb:J1810_par}.

\section{Discussion} \label{sec:dis}
\subsection{Binary evolution and system properties}
PSR~J1810$-$0623 represents a valuable addition to the population of long–orbital-period MSPs with CO-WD companions. With a spin period of 4.55\,ms, an orbital period of 15.4\,days, and an extremely low eccentricity ($e \simeq 1.5 \times 10^{-5}$), the system closely matches the properties expected from an IMXB evolutionary channel. In particular, assuming a pulsar mass of 1.4\,$M_\odot$, the inferred median companion mass of $\sim$0.64\,$M_\odot$ strongly supports a CO-WD classification, consistent with population-synthesis predictions for this formation pathway.

\begin{figure*}[]
    \centering
    \includegraphics[height=7cm,width=18cm]{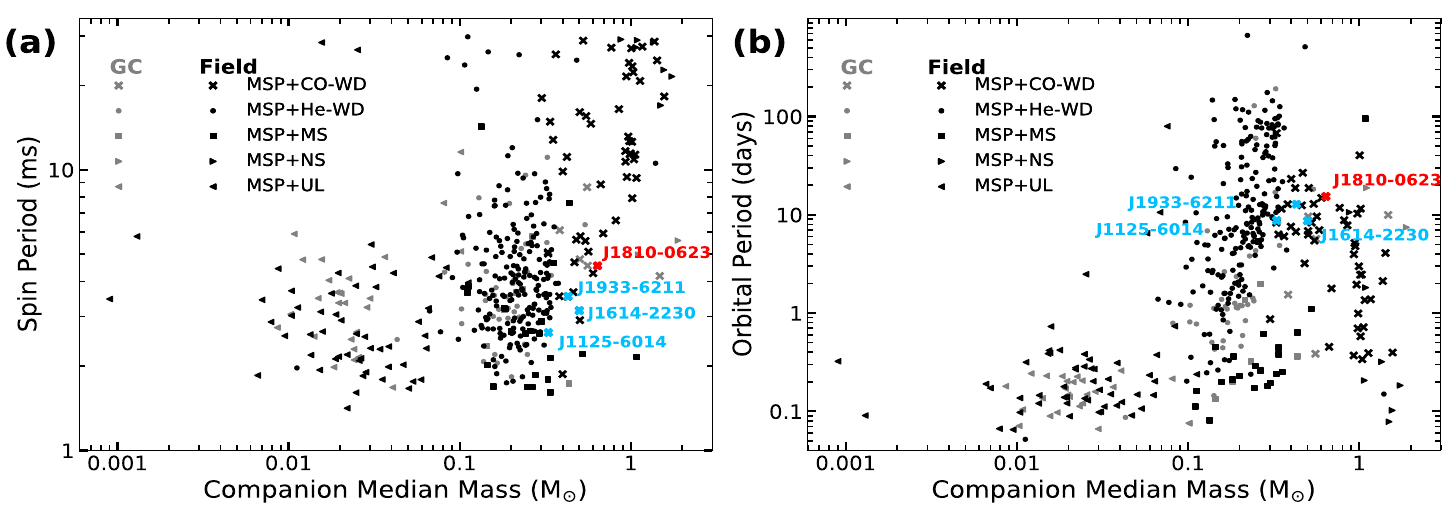}
     \caption{Binary properties of MSPs in the Galactic field and globular clusters (based on the ATNF Pulsar Catalogue$^{\ref{ATNF_catalog}}$). (a) Spin period versus median companion mass. (b) Orbital period versus median companion mass. Different companion types are indicated by distinct symbols: CO white dwarfs (CO-WDs, crosses), He white dwarfs (He-WDs, circles), main-sequence stars (MS, squares), neutron stars (NS, right-pointing triangles), and ultra-light companions (ULs, left-pointing triangles). Systems in the Galactic field are shown in black, while those in globular clusters are shown in light grey. PSR~J1810$-$0623 (this work) is highlighted in red, while PSRs J1125$-$6014, J1614$-$2230, and J1933$-$6211~\cite{Demorest2010, Shamohammadi2023, Geyer2023} are highlighted in red and blue, respectively.} 
    \label{fig:J1810_MassP0Pb}
\end{figure*}

\begin{figure*}[]
    \centering
    \includegraphics[height=10cm,width=16cm]{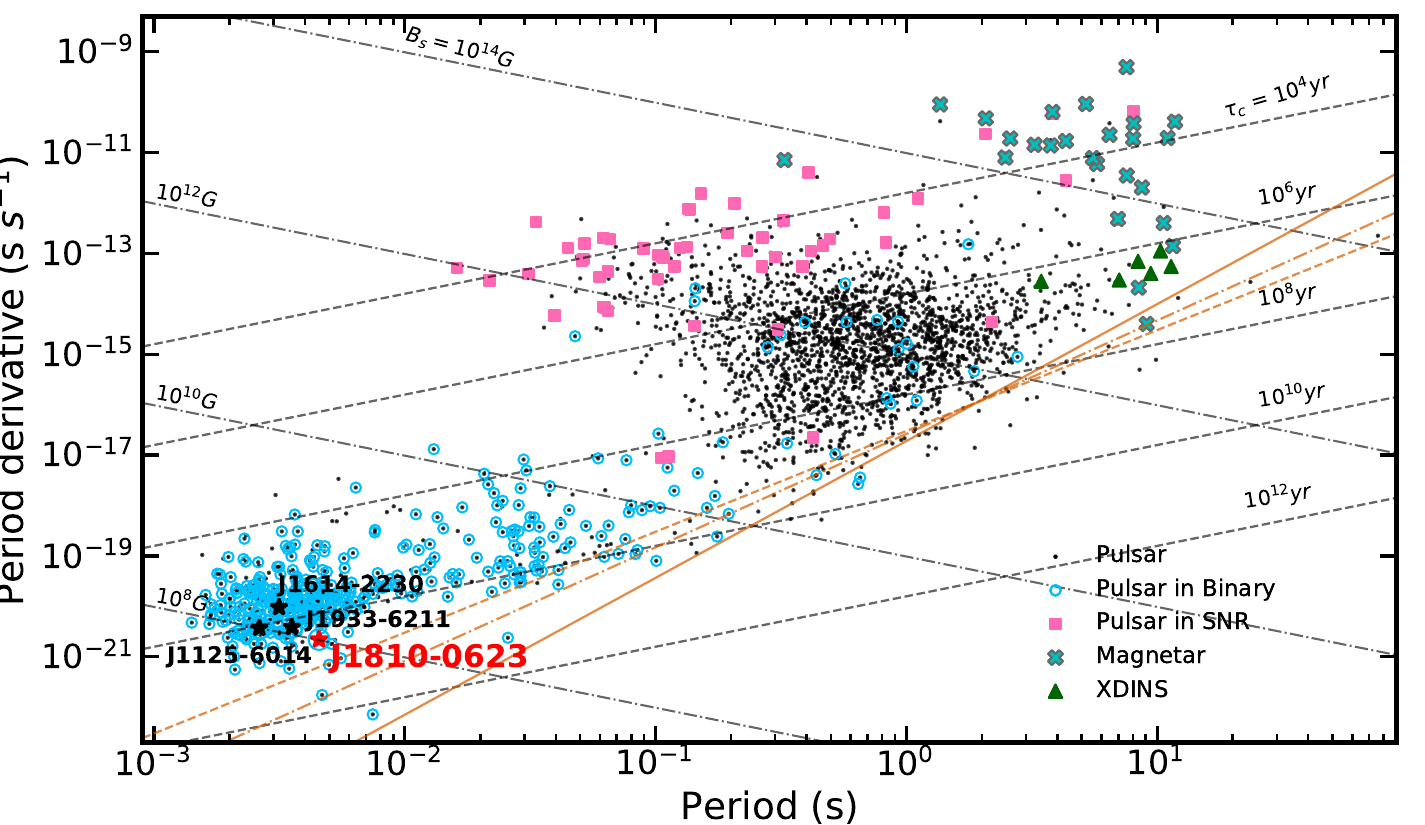}
    \caption{$P$–$\dot{P}$ diagram based on the ATNF pulsar catalogue$^{\ref{ATNF_catalog}}$. Dot-dashed and dashed grey lines denote lines of constant surface magnetic field strength (in Gauss) and characteristic age (in years), respectively. PSR~J1810$+$0623 is marked by a red star outlined in blue, with an inferred characteristic age of $\sim$32\,Gyr. 
    PSRs J1125$-$6014, J1614$-$2230, and J1933$-$6211~\cite{Demorest2010, Shamohammadi2023, Geyer2023} are marked by black stars outlined in blue.} 
    Different pulsar classes are indicated by distinct symbols, as labeled in the legend (e.g., SNR: supernova remnant; XDINS: X-ray dim isolated neutron star). The solid orange curve represents the theoretical pulsar death line for spin-down powered neutron stars with pure dipole magnetic fields~\cite{Chen1993}, while the dashed orange lines correspond to alternative death line predictions incorporating multipole magnetic field components~\cite{Zhang2000}.
\label{fig:J1810_ppdot}
\end{figure*}

Compared with other field MSPs hosting CO–WD companions (see Figure~\ref{fig:J1810_MassP0Pb} and also Figure~\ref{fig:J1810_ppdot}), PSR~J1810$-$0623 occupies a region of parameter space similar to several well studied systems with precisely measured masses via Shapiro delay, including PSRs  J1125$-$6014, J1614$-$2230, and J1933$-$6211~\cite{Demorest2010, Shamohammadi2023, Geyer2023}. These pulsars share several characteristic properties: short spin periods (2.6–3.5 ms), extremely small orbital eccentricities ($e \lesssim 10^{-6}$), relatively massive white dwarf companions ($M_{\rm c} \gtrsim 0.3$–$0.5,M_\odot$), and very small spin period derivatives ($\dot{P} \sim 10^{-21}$ s s$^{-1}$), corresponding to surface magnetic fields of order $10^{8}$ G. Systems with this combination of properties are widely interpreted as products of Case A Roche lobe overflow (RLO; see~\cite{Tauris2011, Tauris2023}), in which mass transfer begins while the donor star is still on the main sequence. This evolutionary channel leads to a prolonged accretion phase that efficiently spins up the neutron star to millisecond periods while strongly suppressing the magnetic field and circularizing the orbit. The observed properties of PSR~J1810$-$0623 are fully consistent with this formation scenario.

From a broader evolutionary perspective, the exceptionally low orbital eccentricity indicates efficient circularization during the mass-transfer phase, as predicted by standard binary evolution models. The pulsar’s position well above theoretical death lines (Figure~\ref{fig:J1810_ppdot}) confirms that it remains comfortably active as a radio pulsar. The characteristic age of $\sim$32\,Gyr inferred from $P/\dot{P}$ is clearly unphysical; this arises from the extremely small observed period derivative ($\dot{P} \sim 2.2 \times 10^{-21}$\,s\,s$^{-1}$) and is a common feature among fully recycled MSPs, where magnetic braking no longer dominates the spin evolution. Future measurements of the intrinsic $\dot{P}$, corrected for Shklovskii and Galactic acceleration effects, may provide a more meaningful estimate of the pulsar’s evolutionary timescale.

The measured rotation measure (RM = $86.6 \pm 0.6$\,rad\,m$^{-2}$) provides an independent probe of the Galactic magnetic field along this line of sight. Comparison with recent Faraday rotation maps \cite{Hutschenreuter2022, Hutschenreuter2024} shows that both the magnitude and sign of the RM are fully consistent with expectations at this location ($l = 22.53^\circ$, $b = +6.11^\circ$). This agreement supports the large-scale Galactic magnetic field configuration inferred from these models and highlights the utility of well-localized pulsars such as PSR~J1810$-$0623 as precise tracers of the Galactic magneto-ionic medium.

\subsection{Prospects of measuring the Shapiro delay}
The relatively high mass of the companion to PSR J1810+0623 makes the prospect of measuring the Shapiro delay enticing. As discussed by Freire \& Wex (2010) \citep{Freire_Wex2010}, for near-circular orbits, the first two harmonics of the Shapiro delay are always absorbed by a re-definition of the classical R{\o}mer delay. For low-inclination orbits, this absoption results in an apparent small eccentricity with an associated longitude of periastron $\omega$ close to $90^\circ$. For PSR J1810+0623, we measure $e = (1.55\pm 0.02) \times 10^{-5}$ with $\omega = (305.5 \pm 0.6) ^\circ$, so this is not the case. In general, only the third ($h_3$) and higher-order harmonics of the Shapiro delay can in fact be measured. We attempted to detect $h_3$ in PSR~J1810$+$0623 using an ELL1H timing model in \texttt{TEMPO}, but found no evidence for it. However, our current data does not cover the pulsar's superior conjunction, and the ToAs available to us are very sparse in the vicinity of it. 

\begin{figure*}[]
    \centering
    \includegraphics[width=0.6\linewidth]{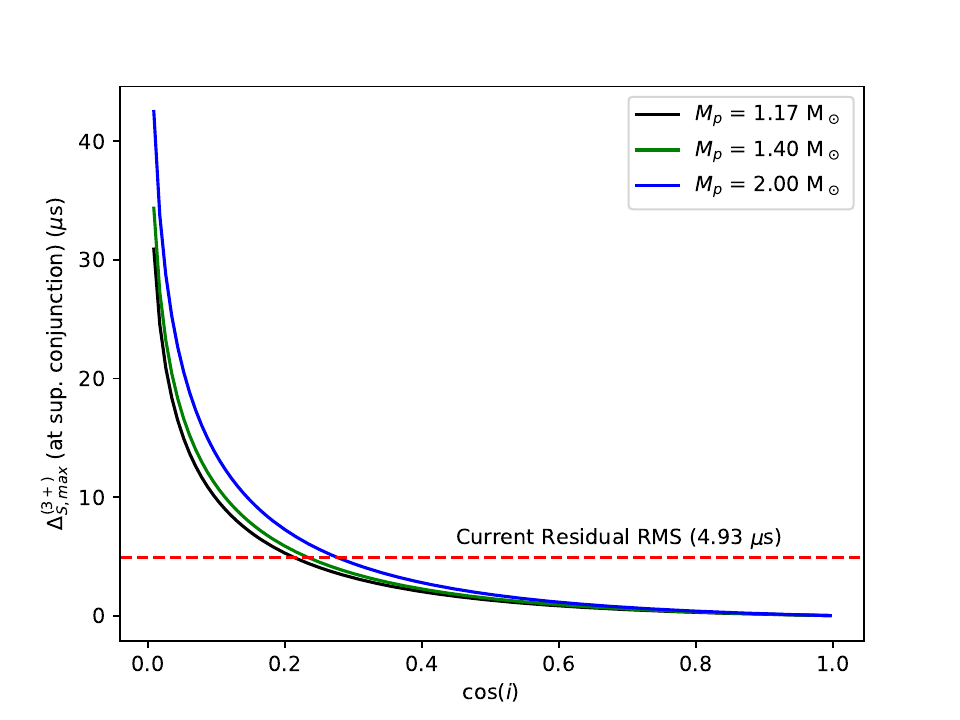  }
    \caption{Predicted maximum amplitude (at superior conjunction) of the Shapiro delay contribution due to the third and higher-order harmonics as a function of the cosine of the inclination, for three different assumed masses for the pulsar. The dashed horizontal line indicates the residual rms of the current timing solution.}
    \label{fig:shapiro_h3+_vs_cosi}
\end{figure*}

To understand whether the effect could potentially be measured in the system, we calculated the peak $\Delta_{\rm S, max}^{(3+)}$ (occurring at the pulsar's superior conjunction) of the contribution of the third and all successive harmonics of the Shapiro delay, as per Eq. (28) of Freire \& Wex (2010), for all the possible orbital inclinations and all plausible combinations of pulsar/companion masses. 
We considered three cases, namely a pulsar mass of 1.17~$M_\odot$,  1.4~$M_\odot$ and 2~$M_\odot$. For each of these cases, we used the mass function to calculate, for all possible orbital inclinations, the companion mass, $M_{\rm c}$, and the maximum value of $\Delta_{\rm S}^{(3+)}$, occurring at the pulsar's superior conjunction, and compared it with our current timing precision. As shown in Figure \ref{fig:shapiro_h3+_vs_cosi}, we find  $\Delta_{\rm S, max}^{(3+)}$ to be equal or larger than our current timing residual rms at inclinations larger than $\sim78$, $\sim76.5$ and $\sim74$ deg for $M_{p} = 1.17, 1.40~{\rm and}\, 2~M_\odot$, respectively, and two times our current timing residual rms at inclinations larger than $\sim84.5$, $\sim83.5$ and $\sim82$ deg for $M_{p} = 1.17, 1.40~{\rm and}\, 2.00~M_\odot$, respectively.

Considering perfectly randomly oriented orbits, these conditions translate into probabilities of 21, 23 and 28 percent for $\Delta_{\rm S, max}^{(3+)} \gtrsim {\rm rms}$ and 10, 11 and 14 percent for $\Delta_{\rm S, max}^{(3+)} \gtrsim 2 \times {\rm \, rms}$. Therefore, the probabilities of measuring a significant  Shapiro delay in PSR~J1810+0623 is high enough to justify future observing campaigns with FAST specifically for the purpose, e.g. with observations made around the pulsar's superior conjunction. The detection of the Shapiro delay would be extremely valuable as it would allow us to constrain the system inclination and, therefore, the individual component masses.

An additional motivation for measuring the Shapiro delay in PSR J1810$-$0623 arises from its likely Case A RLO origin. During such prolonged accretion phases, neutron stars may accrete up to $ \sim 0.3 M_{\odot}$~\cite{Tauris2011}, potentially producing unusually massive neutron stars, as demonstrated by PSR~J1614$-$2230. Precise mass measurements for systems in this evolutionary class are therefore of particular importance, both for constraining accretion efficiency during binary evolution and for probing the neutron star equation of state.

\subsection{Kinematics and origin}
The measured composite proper motion of PSR~J1810$+$0623 ($\mu = 16.77 \pm 7.70$\,mas\,yr$^{-1}$) implies a 2D transverse velocity of $\sim$16 to 80\,km\,s$^{-1}$ for distance estimates between 0.2 and 1.0\,kpc, based on the YMW16 and NE2001 electron density models. This velocity range is fully consistent with that of recycled MSPs in the Galactic field. Given the low Galactic latitude of the source ($b \simeq 6^\circ$), uncertainties in the distance, driven by model dependent dispersion measure contributions, dominate the inferred velocity range. Continued long term timing will further refine the proper motion measurement and allow a more robust assessment of the system’s intrinsic kinematics.

The possibility that PSR~J1810$+$0623 originated in the nearby globular cluster IC~1276 (Pal~7) is worth consideration. The cluster centre lies approximately 50\,arcmin from the current position of the pulsar, well beyond the cluster tidal radius of $\sim$21\,arcmin \cite{Harris1996}. While such a separation makes a bound association unlikely, dynamical ejection scenarios via exchange encounters or tidal interactions could, in principle, place a pulsar outside the nominal tidal boundary. However, several independent lines of evidence argue strongly against this interpretation. The observed dispersion measure of PSR~J1810$+$0623 (24.4\,pc\,cm$^{-3}$) is substantially lower than the $\sim$150 to 180\,pc\,cm$^{-3}$ predicted by both the NE2001 and YMW16 models toward Pal~7 at its distance of $\sim$4.55\,kpc \cite{Baumgardt2018}. In addition, Gaia EDR3 measurements yield an absolute proper motion for Pal~7 of $\mu_{\alpha} = -2.553 \pm 0.026$\,mas\,yr$^{-1}$ and $\mu_{\delta} = -4.568 \pm 0.026$\,mas\,yr$^{-1}$ \cite{Vasiliev2021}, which is inconsistent with the proper motion measured for PSR~J1810$+$0623, even when accounting for uncertainties. Taken together, the dispersion measure and kinematic discrepancies strongly disfavor a physical association with Pal~7.

\section{Summary} \label{sec:sum}
We report the discovery and detailed characterization of PSR~J1810$-$0623, a 4.55-ms pulsar identified with FAST and subsequently studied through follow-up observations with FAST and the GBT. Using a phase-connected timing solution spanning more than 6.5 years, we have precisely determined its spin, astrometric, and binary parameters.

PSR~J1810$-$0623 resides in a 15.4-day binary orbit with an exceptionally low eccentricity of $e \simeq 1.5 \times 10^{-5}$. Assuming a canonical neutron star mass of $1.4\,M_\odot$, the inferred companion mass ($M_{\rm c,\,med} \approx 0.64\,M_\odot$) strongly favors a carbon--oxygen white dwarf, consistent with an intermediate-mass X-ray binary evolutionary origin. Its combination of a short spin period and a long, highly circularized orbit places it among the rare population of MSP+CO-WD systems.

The properties of PSR~J1810$-$0623 closely resemble those of several systems with precisely measured masses via Shapiro delay, including PSRs~J1125$-$6014, J1614$-$2230, and J1933$-$6211. These binaries are widely interpreted as products of Case~A Roche lobe overflow. This evolutionary pathway naturally explains the combination of short spin period, suppressed magnetic field, and highly circularized orbit observed in PSR~J1810$-$0623.

Polarimetric analysis of the FAST data reveals a moderately polarized radio profile and a rotation measure of $86.6 \pm 0.6$\,rad\,m$^{-2}$, providing an additional probe of the Galactic magnetic field along this line of sight. The measured spin-down rate implies an extremely low surface magnetic field and a characteristic age far exceeding the age of the Universe, reflecting the highly recycled nature of the pulsar rather than a physical evolutionary timescale.

The measured proper motion implies a modest transverse velocity, consistent with those of recycled millisecond pulsars in the Galactic field, though uncertainties in the distance estimate and potential kinematic biases warrant continued timing to refine the intrinsic parameters. We find no compelling evidence for an origin in the nearby globular cluster Pal~7, as discrepancies in both dispersion measure and proper motion argue against a physical association.

Overall, PSR~J1810$-$0623 represents a valuable addition to the Galactic millisecond pulsar population, offering a clean laboratory for studying pulsar recycling, binary evolution, and Galactic magneto-ionic structure. Continued high-precision timing, particularly with FAST, may enable future measurements of relativistic effects such as the Shapiro delay, further constraining the system geometry and neutron star mass. Detection of the companion at other wavelengths (e.g., optical or X-ray) could offer additional constraints on its nature and help elucidate the formation history of such binary systems~\cite{Bhalerao2011, Pancrazi2012}.

\subsection*{\begin{center}ACKNOWLEDGMENTS\end{center}}
This work is supported by the National Key R\&D Program of China (2023YFB4503305), the National Natural Science Foundation of China (grant Nos. 12373109, 12588202, 12103069, and 12473042), the Key R\&D of Shandong (2022CXGC020106), and the CAS project No. JZHKYPT-2021-06. D.L. is a New Cornerstone Investigator. A.R. gratefully acknowledges continuing support from the Max-Planck Society. P.W. is supported by NSFC grant No. 12041303, the CAS Youth Interdisciplinary Team, the Youth Innovation Promotion Association CAS (id. 2021055), and the Cultivation Project for FAST Scientific Payoff and Research Achievement of CAMS-CAS. This work utilized data from FAST and GBT, with data processing, conducted primarily on the facility at  Qilu Normal University and Jinan Institute of Supercomputing Technology, employed the "Pulsar and Fast Radio Burst Search and Analysis Platform" and its integrated search algorithms—both developed under the 2023YFB4503305 Project.
We thank Paulo Freire and Alessandro Corongiu for useful discussions.\\

{\bf Conflict of interest} The authors declare that they have no conflict of interest.

\bibliographystyle{scpma-zycai} 
\bibliography{ms}

\end{multicols}
\end{document}